%% file: main.tex
\documentclass[10pt,journal]{IEEEtran}

\usepackage{amsmath,amsfonts}
\usepackage{algorithmic}
\usepackage{algorithm}
\usepackage{array}
\usepackage[caption=false,font=normalsize,labelfont=sf,textfont=sf]{subfig}
\usepackage{textcomp}
\usepackage{stfloats}
\usepackage{url}
\usepackage{verbatim}
\usepackage{graphicx}
\usepackage{cite}

\usepackage{tabularx}
\usepackage{booktabs}
\usepackage[normalem]{ulem}
\useunder{\uline}{\ul}{}
\usepackage{longtable}
\usepackage{adjustbox}
\usepackage{CJKutf8}
\usepackage[utf8]{inputenc}
\usepackage{multirow}
\usepackage{enumitem}
\usepackage{xcolor}
\usepackage{microtype}

\usepackage{tikz}
\usetikzlibrary{arrows.meta,positioning,calc}
\begin{document}
\begin{CJK}{UTF8}{gbsn}
\title{Can We Rely on Language Models to Orchestrate Services across the Edge–Cloud Continuum?}

\author{
Zhenghao Wu,
Ruihai Dong,
Reza Farahani,
Hamed Haddadi,
Omer Rana,
and Hadi Tabatabaee Malazi 
\thanks{Zhenghao Wu, Ruihai Dong, and Hadi Tabatabaee~Malazi are with the
School of Computer Science, University College Dublin, Ireland.
E-mail: Zhenghao.Wu@ucdconnect.ie, \{Ruihai.Dong, Hadi.TabatabaeeMalazi\}@ucd.ie}
\thanks{Reza Farahani is with the Distributed Systems Group (DSG),
TU Wien, Austria.
E-mail: R.Farahani@dsg.tuwien.ac.at}
\thanks{Hamed Haddadi is with the Department of Computer Science,
University of Oxford, UK.
E-mail: hamed.haddadi@cs.ox.ac.uk}

\thanks{Omer Rana is with the School of Computer Science and Informatics,
Cardiff University, UK.
E-mail: RanaOF@cardiff.ac.uk}
\thanks{\textit{(Corresponding author: Hadi Tabatabaee Malazi.)}}%
}

\maketitle

\input{sections/000-abstract}

\section{Introduction}
\label{section/introduction}
\input{sections/100-introduction}

\section{Background}
\label{section/background}
\input{sections/300-background}

\section{Survey Methodology}
\label{section/survey-methodology}

\input{sections/400-survey_method}

\section{Orchestration Decisions}
\label{section/rq1}
\input{sections/500-rq1}

\section{LM Roles, Authority, and Assurance}
\label{section/rq2}
\input{sections/600-rq2}

\section{Evaluation Evidence}
\label{section/rq3}
\input{sections/700-rq3}

\section{Future Research Directions}
\label{section/future}
\input{sections/800-future}

\section{Conclusion}
\label{section/conclusion}
\input{sections/900-conclusion}

\footnotesize
\bibliographystyle{IEEEtran}
\bibliography{literatures,extended}

\clearpage
\normalsize

\onecolumn

\appendix
\label{section/apdx}
\input{sections/1099-supp}

\end{CJK}
\end{document}

%% file: sections/000-abstract.tex
\begin{abstract}
Service orchestration across the edge--cloud continuum increasingly uses language models (LMs) to translate intents and operational evidence into scheduling, configuration, diagnosis, and remediation decisions. Yet, useful model outputs do not by themselves establish dependable orchestration: reliability also depends on where the LM enters the decision process, what authority it is granted, how actions are constrained, and what system-level evidence supports the claimed effect. Existing surveys provide limited synthesis of the links between continuum-specific orchestration decisions, LM responsibility, operational authority, assurance controls, and supporting empirical evidence. We systematically review 39 primary studies and develop a five-dimensional analytical framework for connecting system context, orchestration decisions, LM agency, operational safeguards, and evaluation evidence. The synthesis reveals predominantly hybrid designs in which LMs support interpretation, candidate generation, and mitigation planning, while downstream components commonly retain feasibility enforcement and execution authority. Bounded closed-loop remediation is emerging, but continuum-specific constraints remain unevenly represented in decision formulations, and production-scale evidence across heterogeneous edge--cloud environments remains limited. These findings suggest that reliance on LM-enabled orchestration depends on aligning decision constraints, operational authority, safeguards, and evaluation evidence with the claimed system effect.
\end{abstract}

%% file: sections/100-introduction.tex
Delivering dependable services across the edge--cloud continuum requires coordinated decisions about where workloads execute, how resources are allocated, and how services adapt at runtime. These decisions are coupled: a placement that satisfies a latency objective can become unsuitable as workload or connectivity changes, while a local adaptation may affect dependent services elsewhere in the continuum \cite{akbari_intentcontinuum_2025}. Therefore, orchestration must translate high-level objectives into operational changes while considering heterogeneous resources, distributed state, and changing constraints. This translation layer provides a natural entry point for language models (LMs) in orchestration.

LMs can interpret natural-language objectives and heterogeneous operational data, and they can also reason, plan, and interact with managed systems~\cite{akbari_intentcontinuum_2025,shetty_building_2024,avgerinos_arm_2026}. Once their outputs influence orchestration, model capability alone is insufficient: a useful recommendation may still be infeasible, unauthorized, unexecuted, or unverified. The question therefore shifts from whether LMs can generate plausible orchestration outputs to when LM-enabled service orchestration can be relied upon, which depends on the supported decision, the authority and assurance boundary around action, and the empirical evidence behind the claimed system effect.

Existing surveys cover closely related domains, including continuum orchestration, LM-enabled network and service management, AIOps, and agentic operations. Kokkonen et al.~\cite{Kokkonen2023} provide a broad account of orchestration across the device--edge--cloud continuum, emphasizing autonomy and intelligence in continuum orchestration. Boateng et al.~\cite{Boateng2026} survey LLM-enabled communication, network, and service management across mobile, vehicular, cloud-based, and fog/edge-based network environments. Zhang et al.~\cite{Zhang2025AIOps} examine LLM-based AIOps across data sources, tasks, methods, and evaluation, whereas Bilal et al.~\cite{Bilal2026} analyze agentic NetOps and AIOps through architectures, autonomy, operational control, assurance, evaluation, safety, and security.

These surveys organize the literature primarily by continuum paradigm, network or AIOps task, LM application, or operational agent. Although they examine autonomy, assurance, and evaluation, the orchestration decision itself is not their primary unit for relating continuum-specific constraints to LM responsibility, downstream authority, and empirical evidence. We adopt the LM-enabled service-orchestration decision as the unit of comparison, tracing managed state, constraints, and feasible actions through LM participation, downstream control, and evaluated outcomes. This distinguishes deployment context from decision-relevant continuum constraints and reveals where empirical evaluation falls short of the claimed orchestration capability. Table~\ref{tab:related-surveys} summarizes the comparison.

\input{table/table-rw-compare}

This framing leads to three research questions. 
\begin{itemize}
    \item \textbf{RQ1: Which orchestration decisions do LMs support?} This question identifies the orchestration problems, decision variables, and continuum constraints represented in the literature.
    \item \textbf{RQ2: How are LM responsibility, operational authority, and assurance controls distributed across the orchestration decision path?} This question traces LM decision roles and downstream handoffs, including which components validate, constrain, authorize, or execute consequential actions. 
    \item \textbf{RQ3: What empirical evidence supports the claimed effects and operational maturity of LM-enabled orchestration?} This question examines the evaluated endpoint and whether it supports the decision or system outcome implied by the claim.
\end{itemize}

We answer these questions through a systematic mapping of 39 primary studies, organized using the five-dimensional SPACE framework: System Context, Problem and Decision, LM Agency, Control and Assurance, and Evidence and Maturity. The synthesis reveals predominantly hybrid designs in which LMs support semantic interpretation, candidate generation, and mitigation planning, while downstream components commonly retain feasibility enforcement, authorization, and privileged execution. Emerging closed-loop remediation remains bounded by action, resource, and execution scopes. Evaluations span artifact validity, decision quality, executed interventions, and verified recovery, supporting different levels of claimed orchestration capability. 

The survey makes the following contributions:

\begin{enumerate}
\item  We use the LM-enabled service-orchestration decision as our unit of analysis, linking formulation, managed state, LM participation, operational authority, assurance boundaries, and evidence supporting claimed effects. 

\item Applying this perspective, we show when edge--cloud continuum conditions merely provide deployment context and when they actively shape orchestration decisions, examining how managed states, decision constraints, feasible actions, and state transitions differ across resource/workload, configuration/policy, resilience/operations, and workflow orchestration.

\item We distinguish LM responsibility from operational authority
by tracing who selects, authorizes, and executes consequential
actions, whether independent safeguards constrain them, and
how post-action recovery is handled. 

\item We assess whether reported evaluations support claimed orchestration capabilities, distinguishing artifact validity, decision quality, executed interventions, and verified recovery under stated operating conditions.

\item We identify security and sustainability gaps in LM-enabled orchestration, including the distinction between security-oriented uses and the security of the LM-mediated orchestration process itself, incomplete accounting of LM and control overheads, and limited evidence on carbon intensity and hardware lifetime.
\end{enumerate}

The remainder of this survey presents the background (\S \ref{section/background}) and methodology (\S \ref{section/survey-methodology}), followed by the three research questions on orchestration decisions (\S \ref{section/rq1}), LM agency and assurance (\S \ref{section/rq2}), and evaluation evidence (\S \ref{section/rq3}), and finally the future research directions (\S \ref{section/future}) and conclusion (\S \ref{section/conclusion}). \vspace{-0.15in}

%% file: table/table-rw-compare.tex
\begin{table*}[t]
\centering
%\caption{Positioning against closely related surveys. A checkmark denotes a dedicated analytical dimension, $\triangle$ denotes substantive but partial or task-specific treatment, and --- denotes no substantive analytical treatment.}
\caption{Positioning of this survey against closely related surveys.}
\label{tab:related-surveys}

\scriptsize
\setlength{\tabcolsep}{2.6pt}
\renewcommand{\arraystretch}{1.10}

\begin{tabular}{@{}lcccccccccc@{}}
\toprule
&
\multicolumn{1}{c}{\textbf{Context}}
&
\multicolumn{4}{c}{\textbf{RQ1: Orchestration decisions}}
&
\multicolumn{2}{c}{\textbf{RQ2: LM integration}}
&
\multicolumn{1}{c}{\textbf{RQ3}}
&
\multicolumn{2}{c}{\textbf{Cross-cutting}}
\\
\cmidrule(lr){2-2}
\cmidrule(lr){3-6}
\cmidrule(lr){7-8}
\cmidrule(lr){9-9}
\cmidrule(l){10-11}

\textbf{Survey}
&
\shortstack{\textbf{Edge--cloud}\\\textbf{continuum}}
&
\shortstack{\textbf{Resource /}\\\textbf{workload}}
&
\shortstack{\textbf{Config. /}\\\textbf{policy}}
&
\shortstack{\textbf{Resilience /}\\\textbf{operations}}
&
\shortstack{\textbf{External}\\\textbf{workflow}}
&
\shortstack{\textbf{Roles /}\\\textbf{authority}}
&
\shortstack{\textbf{Operational}\\\textbf{assurance}}
&
\shortstack{\textbf{Evaluation}\\\textbf{evidence}}
&
\textbf{Security}
&
\shortstack{\textbf{Sustainability}\\\textbf{/ SDG}}
\\
\midrule

Kokkonen et al.~\cite{Kokkonen2023}
& $\checkmark$
& $\checkmark$
& $\triangle$
& $\triangle$
& $\triangle$
& ---
& $\triangle$
& ---
& $\triangle$
& $\triangle$
\\

Boateng et al.~\cite{Boateng2026}
& $\checkmark$
& $\checkmark$
& $\checkmark$
& $\checkmark$
& $\triangle$
& $\triangle$
& $\triangle$
& $\triangle$
& $\checkmark$
& ---
\\

Zhang et al.~\cite{Zhang2025AIOps}
& $\triangle$
& ---
& $\triangle$
& $\checkmark$
& ---
& $\triangle$
& ---
& $\checkmark$
& ---
& ---
\\

Bilal et al.~\cite{Bilal2026}
& $\triangle$
& ---
& $\checkmark$
& $\checkmark$
& ---
& $\checkmark$
& $\checkmark$
& $\checkmark$
& $\checkmark$
& ---
\\

\textbf{This survey}
& $\checkmark$
& $\checkmark$
& $\checkmark$
& $\checkmark$
& $\checkmark$
& $\checkmark$
& $\checkmark$
& $\checkmark$
& $\checkmark$
& $\checkmark$
\\

\bottomrule
\end{tabular}

\vspace{2pt}
\begin{minipage}{0.98\textwidth}
\scriptsize
\emph{Legend:} $\checkmark$ = used as a dedicated analytical dimension across the survey;
$\triangle$ = discussed substantively but only within selected tasks, examples, or subsections;

--- = not treated as a substantive analytical dimension.

%old version \emph{Note:} ``External workflow'' refers to a workflow whose execution state is itself managed by the orchestration system, rather than an internal multi-agent or reasoning workflow.
\emph{Note:} ``External workflow'' means a service or deployment workflow managed by the orchestration system, not an internal multi-agent or reasoning workflow.
\end{minipage}
\vspace{-0.2in}
\end{table*} 

%% file: sections/300-background.tex
\begin{figure}[!t]
    \centering
    \includegraphics[width=\linewidth]{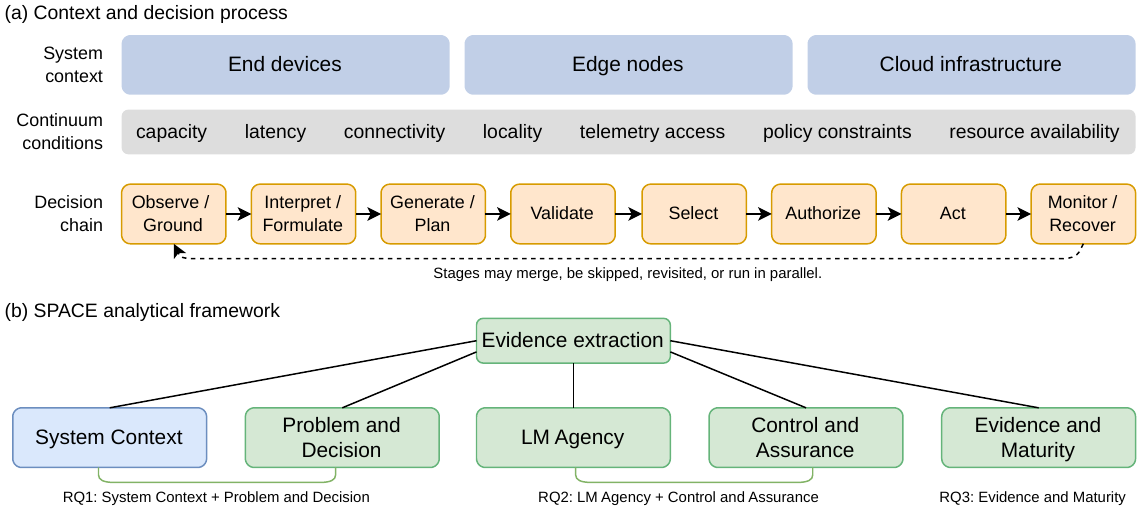}
    \caption{Decision chain, SPACE, and their mapping to the research questions.}
%    \caption{Unified analytical framework linking the orchestration decision chain and SPACE dimensions to the three research questions.}
    \label{fig:space-taxonomy}
    \vspace{-0.15in}
\end{figure}

Edge--cloud orchestration coordinates heterogeneous services and infrastructure through decisions that affect workload placement, resource allocation, configuration, policy, workflow progression, and recovery state. Figure~\ref{fig:space-taxonomy} summarizes the decision chain and SPACE dimensions this survey uses to compare decision context, LM responsibility, operational authority, assurance mechanisms, and supporting evidence. \vspace{-0.1in}

\subsection{Orchestration Scope}
The edge–cloud continuum joins execution environments that differ in compute capacity, latency, connectivity, locality, observability, administrative control, and exposure to changing workloads. Orchestration manages services, workloads, resources, configurations, policies, workflow states, and failure-recovery states across these environments; therefore, a feasible placement or adaptation can shift as network and resource conditions evolve \cite{akbari_intentcontinuum_2025}. 

We use LM as an umbrella term encompassing both large and compact models used in edge--cloud orchestration, whether deployed locally, at the edge, in the cloud, or accessed as an external service. Rather than treating the LM as an end-to-end system replacement, we position it as a participant in a broader orchestration pipeline. Our scope includes LM outputs that either materially shape orchestration choices or supply decision-relevant information used to formulate, select, validate, or revise such choices. This survey excludes generic model deployment and isolated monitoring or summarization without an identifiable orchestration role. \vspace{-0.1in}

\subsection{Decision Chain}
We conceptualize orchestration as a unified functional chain: Observe/Ground $\rightarrow$ Interpret/Formulate $\rightarrow$ Generate/Plan $\rightarrow$ Validate $\rightarrow$ Select $\rightarrow$ Authorize $\rightarrow$ Act $\rightarrow$ Monitor/Recover. Although implementations may merge, skip, revisit, or parallelize these steps, the chain allows us to isolate distinctions that architectural labels often obscure.

LM participation is decision-enabling when the model produces evidence, interpretations, objectives, explanations, or candidate sets for selection by a human operator or non-LM decision component. It is decision-direct when the model specifies an action, plan, policy, or critical parameter as the proposed orchestration decision. We then distinguish this decision responsibility from operational authority by recording whether the output is selected for action, authorized, executed, or verified after execution. \vspace{-0.1in}

\subsection{SPACE Framework}
\label{sec:space}

To provide a common basis for comparing heterogeneous implementations, SPACE organizes the synthesis along five dimensions: System Context, Problem and Decision, LM Agency, Control and Assurance, and Evidence and Maturity. These dimensions specify where decisions arise, which system states can change, how responsibility is distributed, which controls constrain progression toward action, and what empirical evidence supports each claim.

The three research questions map directly onto this framework. RQ1 characterizes the system contexts and orchestration decision problems captured by System Context and Problem and Decision. RQ2 examines how responsibility, authority, and safeguards are distributed along the decision path, linking LM Agency with Control and Assurance. RQ3 evaluates Evidence and Maturity by asking whether reported environments, empirical endpoints, comparators, and outcomes support the claimed orchestration effects. \vspace{-0.15in}

%% file: sections/400-survey_method.tex
The survey followed three stages: planning, conduct, and reporting, covering search design, study identification and screening, and evidence extraction and synthesis. Figure~\ref{fig:research-method} summarizes the workflow, inspired by PRISMA 2020~\cite{Pagen71}.

\input{sections/flowchart}

\subsection{Study Selection}

We used Semantic Scholar as the primary retrieval source and considered publications from 2023 through July 2026. Venue and source metadata were cross-checked against publisher records, including IEEE Xplore and the ACM Digital Library. The search used eight primary query families covering \textit{orchestration, scheduling, placement/offloading, scaling, resource management, remediation, intent/policy translation, and zero-touch operations}, supplemented by two query families for \textit{execution safety} and \textit{diagnosis-to-remediation handoffs}. Appendix~\ref{app:reproducibility} reports the search date, query families, screening settings, decision labels, and operational selection criteria.

The search returned 1,808 records. Deduplication reduced this set to 1,443, and deterministic desk screening removed records outside the publication window, records without a language-model component, records without a systems or operations context, and records clearly unrelated to orchestration, remediation, configuration, placement, resource management, or service management, leaving 1,200 records.

We then screened the remaining 1,200 records using two separately prompted assessments with the same model, \texttt{deepseek-v4-flash (preview)}, with temperature 0 and structured JSON output. Each assessment received the available bibliographic metadata, title, and abstract and applied the same scope criteria; the second assessment did not receive the first assessment’s decision. The two assessments agreed that 449 records were potentially eligible and disagreed on eligibility for 85 records (7.1\% of the screened set). Records advanced to full-text assessment only when both assessments judged them potentially eligible with confidence $\geq 0.8$, supplied valid textual evidence supporting the LM role and orchestration relevance, and raised no unresolved publication-type ambiguity. This joint advancement rule produced 421 candidates for full-text assessment.

The 421 candidates were then manually assessed at full text against the stated scope criteria. Eligibility was determined from the reported operational role of the LM and supporting full-text evidence rather than keyword occurrence alone. We included state-changing orchestration decisions as well as diagnosis and decision-support functions when they explicitly informed, constrained, or evaluated incident management, remediation, configuration, resource management, or another orchestration process. We excluded generic monitoring, summarization, model deployment, and language tasks without an identifiable orchestration role.

We then applied forward and backward citation snowballing to eligible studies. Newly identified records were deduplicated and screened using the same scope criteria. Full-text assessment and snowballing produced 41 candidate studies; two non-empirical magazine articles were excluded, leaving 39 primary studies.

\subsection{Evidence Extraction and Synthesis}

For each included study, we extracted a common set of fields describing the orchestration problem and system context, the LM's operational role and downstream handoffs, decision authority and safeguards, and the evaluation setting, endpoints, comparators, and reported outcomes. A common extraction structure was necessary because the reviewed studies use different terminology and often combine several decision stages within the same system. When a study included materially different decision paths, execution modes, or approval conditions, we retained them separately rather than collapsing them into a single study-level description.

We used an LM-assisted process to identify relevant passages in the full text and populate the corresponding evidence fields. These records served as an indexing and comparison aid for the qualitative synthesis, not as an independently analyzed quantitative dataset. Evidence used in the manuscript, figures, and comparative tables was manually checked against the corresponding source material, and ambiguous classifications were resolved through author discussion. The extracted evidence was compared across studies to identify recurring decision patterns, responsibility boundaries, assurance mechanisms, evaluation practices, and gaps in reported evidence. 

%% file: sections/flowchart.tex
\begin{figure}[t]
\centering

\begin{tikzpicture}[
    font=\sffamily\scriptsize,
    >=Latex,
    node distance=2.4mm,
    flow/.style={
        ->,
        line width=0.55pt,
        draw=black
    },
    dottedflow/.style={
        ->,
        line width=0.55pt,
        draw=black,
        densely dotted
    },
    box/.style={
        draw=black,
        fill=gray!10,
        line width=0.55pt,
        minimum width=3.85cm,
        minimum height=0.62cm,
        text width=3.55cm,
        align=center,
        inner sep=1.5pt
    },
    sidebox/.style={
        draw=black,
        fill=white,
        line width=0.55pt,
        minimum width=2.70cm,
        minimum height=0.62cm,
        text width=2.50cm,
        align=center,
        inner sep=1.5pt
    },
    phasebar/.style={
        draw=gray!65,
        fill=gray!10,
        line width=0.45pt,
        inner xsep=4pt,
        inner ysep=2pt
    }
]

% ============================================================
% PLAN
% ============================================================

\node[box] (rq)
    {Research Questions\\
    RQ1--RQ3};

\node[box, below=of rq] (search)
    {Search Strategy\\
     8 task + 2 supplementary queries};

\draw[flow] (rq.south) -- (search.north);

% ============================================================
% CONDUCT
% ============================================================

\node[box, below=4.0mm of search] (retrieval)
    {Database Retrieval\\
     1,808 records};

\node[box, below=of retrieval] (dedup)
    {Deduplication\\
     1,443 retained (-365)};

\node[box, below=of dedup] (desk)
    {Desk Screening\\
     1,200 retained (-243)};

\node[box, below=of desk] (llm)
    {LLM-assisted Screening\\
     421 candidates (-779)};

\node[box, below=of llm] (fulltext)
    {Full-text Eligibility\\
     Assessment};

\node[box, below=of fulltext] (candidate)
    {Full-text assessment\\
 + snowballing\\
 41 candidate studies};

\draw[flow] (search.south) -- (retrieval.north);
\draw[flow] (retrieval.south) -- (dedup.north);
\draw[flow] (dedup.south) -- (desk.north);
\draw[flow] (desk.south) -- (llm.north);
\draw[flow] (llm.south) -- (fulltext.north);
\draw[flow] (fulltext.south) -- (candidate.north);

% ============================================================
% SNOWBALLING
% ============================================================

% First side cell: citation chasing
\node[
    sidebox,
    right=5.5mm of fulltext
] (snowball)
    {Forward/backward\\
     Snowballing};

\node[
    sidebox,
    right=5.5mm of candidate
] (snowcheck)
    {Deduplication\\
     + Screening};

% Full-text review -> snowballing
\draw[flow]
    (fulltext.east) -- (snowball.west);

% Snowballing -> deduplication/screening
\draw[flow]
    (snowball.south) -- (snowcheck.north);

\draw[flow]
    (snowcheck.west) -- (candidate.east);

% ============================================================
% REPORT
% ============================================================

\node[
    box,
    below=4.8mm of candidate
] (included)
    {Included Studies\\
     39 primary studies};

\node[
    box,
    below=of included
] (synthesis)
    {Structured Evidence Extraction\\
     + Comparative Synthesis};

% Main transition: 41 -> 39
\draw[flow]
    (candidate.south) -- (included.north);

\draw[flow]
    (included.south) -- (synthesis.north);

% ============================================================
% EXCLUSION
% ============================================================

% Third side cell
\node[
    sidebox,
    below=2.4mm of snowcheck
] (excluded)
    {Excluded\\
     2 magazine articles};

% Branch from the 41 -> 39 transition
\coordinate (excludePoint)
    at ($(candidate.south)!0.5!(included.north)$);

% Move outside the main column first, then route to the box
\coordinate (excludeBranch)
    at ([xshift=2.5mm]candidate.east |- excludePoint);

\draw[flow]
    (excludePoint)
    -- (excludeBranch)
    |- (excluded.west);

% ============================================================
% PHASE BARS
% ============================================================

% Horizontal positions relative to the left edge of main cells
\def\phaseLeft{-11.0mm}
\def\phaseRight{-7.0mm}
\def\phaseCenter{-9.0mm}

% ------------------------------------------------------------
% PLAN
% ------------------------------------------------------------

\path[
    draw=gray!65,
    fill=gray!10,
    line width=0.45pt
]
    ([xshift=\phaseLeft]rq.north west)
    rectangle
    ([xshift=\phaseRight]search.south west);

\node[
    rotate=90,
    font=\sffamily\scriptsize\bfseries
] at
    ($([xshift=\phaseCenter]rq.north west)!
      0.5!
      ([xshift=\phaseCenter]search.south west)$)
    {Planning};

% ------------------------------------------------------------
% CONDUCT
% ------------------------------------------------------------

\path[
    draw=gray!65,
    fill=gray!10,
    line width=0.45pt
]
    ([xshift=\phaseLeft]retrieval.north west)
    rectangle
    ([xshift=\phaseRight]candidate.south west);

\node[
    rotate=90,
    font=\sffamily\scriptsize\bfseries
] at
    ($([xshift=\phaseCenter]retrieval.north west)!
      0.5!
      ([xshift=\phaseCenter]candidate.south west)$)
    {Conduct};

% ------------------------------------------------------------
% REPORT
% ------------------------------------------------------------

\path[
    draw=gray!65,
    fill=gray!10,
    line width=0.45pt
]
    ([xshift=\phaseLeft]included.north west)
    rectangle
    ([xshift=\phaseRight]synthesis.south west);

\node[
    rotate=90,
    font=\sffamily\scriptsize\bfseries
] at
    ($([xshift=\phaseCenter]included.north west)!
      0.5!
      ([xshift=\phaseCenter]synthesis.south west)$)
    {Reporting};

\end{tikzpicture}

\caption{Study identification, screening, inclusion, and synthesis workflow.}
%Study identification, screening, inclusion, and synthesis workflow. The dotted path denotes supplementary forward and backward citation snowballing.
\label{fig:research-method}
\vspace{-0.15in}
\end{figure}
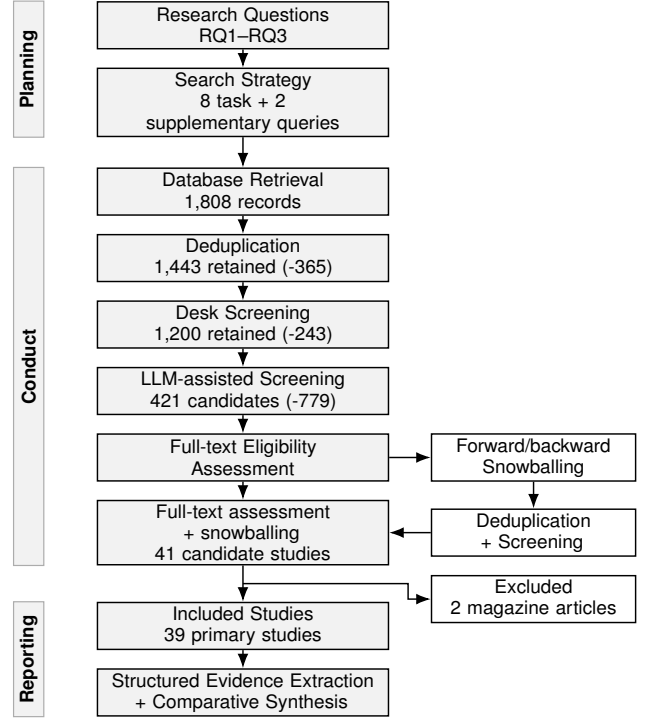

%% file: sections/500-rq1.tex
Edge--cloud orchestration problems differ by the managed state they affect, how changes to that state are formulated as decisions, and which operational constraints bound feasible action. RQ1 organizes this decision space into four recurring domains: resource and workload scheduling, configuration and policy generation, resilience and operations, and workflow orchestration (Table~\ref{tab:decision-landscape}). Across these domains, we compare the managed entity, decision variable, constraints, and resulting state transition. We also distinguish how directly the LM output shapes the transition, from structuring objectives and candidate spaces to proposing operational actions.

\subsection{Decision Problems}

\begin{figure*}
    \centering
    \includegraphics[width=0.8\linewidth]{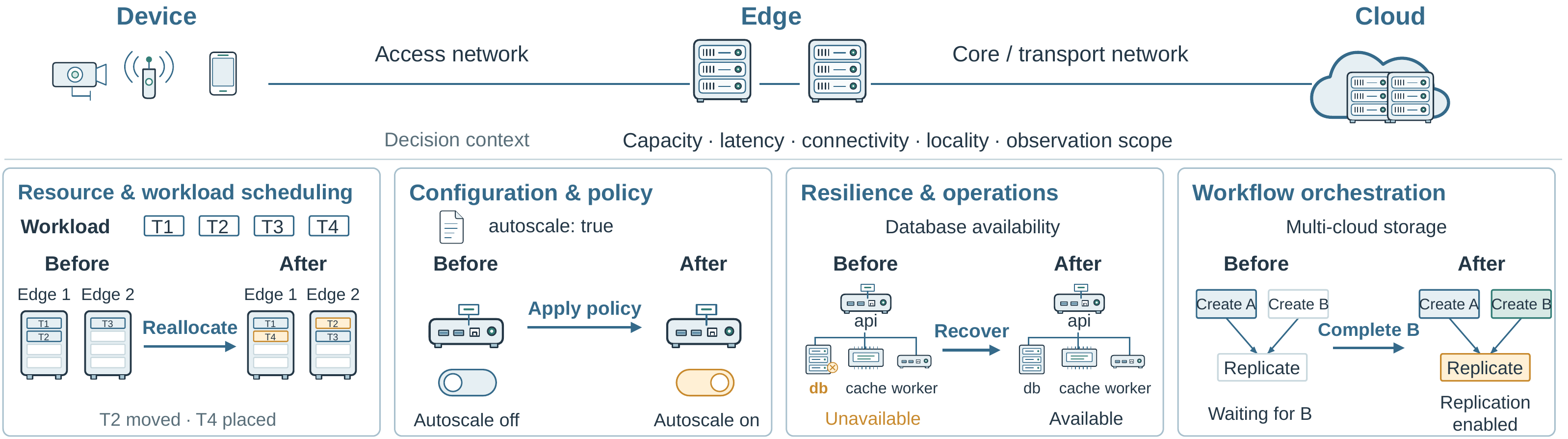}
    \caption{Orchestration decision families and their managed state transitions under edge--cloud constraints.}
    \label{fig:rq1}
    \vspace{-0.15in}
\end{figure*}

\input{table/table-decision-landscape}

\textbf{1) Resource and workload orchestration} assigns tasks to computing resources. LMs enter these allocation problems by translating placement preferences, generating candidate assignments, and constructing optimization rewards. At the representational end, LMs translate semantic placement requirements into scheduler-readable preferences or objectives~\cite{sliwko_cluster_2026}. A second pattern uses the LM to shape the search process. A lightweight LM~\cite{tang_llm-assisted_2025} can generate scheduling candidates from task and virtual-machine states for selection by a Deep Q-Network. L2D2~\cite{lv_large_2026} generates and refines reward functions from environmental feedback for D3QN optimization. A further pattern brings the LM closer to the allocation choice by generating candidate schedules or workload--resource mappings from runtime state. LLMSched~\cite{ding_llm-driven_2026} combines cluster utilization, task dependencies, and service-level requirements to generate candidate assignments, then refines them via Integer Linear Programming under resource constraints.

These problem formulations reveal a recurring division of responsibility between semantic reasoning and constrained optimization. LM outputs can progressively move from describing how an allocation should be evaluated to proposing where a workload should be executed. However, capacity, dependency, and admissibility constraints commonly remain external to the model. Resource orchestration thus spans a spectrum of LM influence over allocation decisions rather than a binary distinction between advisory and direct scheduling. 

\textbf{2) Configuration and policy generation} differs primarily in how far LM output travels toward operational use. At one end, LMs analyze existing infrastructure definitions to identify vulnerabilities or misconfigurations and provide remediation guidance to human-facing development workflows~\cite{toprani_llm_2025,ghorab_towards_2025}. A second group generates explicit, machine-readable artifacts, such as Kubernetes manifests or infrastructure configurations, that specify a target state but still undergo refinement or validation before operational use~\cite{kratzke_dont_2024,aabirrouche_towards_2026}. %Across these studies, the LM moves from describing what should be changed to expressing the intended configuration itself.

Further along this spectrum, generated configurations and policies are integrated with downstream system components. In federated learning, LM-produced task configurations and model architectures are consumed by the parameter server to run the distributed training process~\cite{mawela_web-based_2025}. In transient IoT environments, generated access-control policies feed the policy and auto-configuration mechanisms that determine device permissions~\cite{shastri_llm-driven_2025}, while cloud-native security policy generation produces service-mesh policies intended for direct enforcement~\cite{chen_end--end_2025}. Across these studies, the principal distinction lies in how directly the LM output is coupled to the managed state. Outputs range from recommendations that require human interpretation, through machine-readable artifacts that specify the intended configuration, to artifacts directly consumed by downstream orchestration or enforcement components. Validation and refinement cut across these forms rather than form a separate stage; they may constrain artifacts before human adoption or automated use.

\textbf{3) Resilience and operations} spans a longer decision path, from fault interpretation to system recovery. A substantial group of systems concentrates on narrowing the recovery problem by detecting anomalies, classifying faults, localizing likely causes, reconstructing failure paths, or synthesizing heterogeneous diagnostic evidence~\cite{chen_automatic_2024,li_magmascope_2026,li_coca_nodate,zhang_tamofine-grained_2025,liu_lineadapter_2025,pedroso_anomaly_2025}. These outputs can materially shape incident response, but they remain upstream of the state-changing decision.

Other studies move from diagnosis toward the recovery action itself. Mitigation-oriented approaches generate recovery plans or repair artifacts that can subsequently be checked or executed~\cite{sarda_leveraging_2026,wiesinger_leveraging_2026}. Closed-loop formulations~\cite{avgerinos_arm_2026, zhang_e2e-reme_2026} go further by connecting diagnosis to corrective action and then observing whether the intervention restored the required system state. Resilience-oriented systems therefore progress from diagnostic evidence through mitigation formulation to executable, verified recovery. This progression is more informative than grouping all RCA and remediation systems under a single automation label.

\textbf{4) Workflow orchestration} appears in fewer studies and should be defined narrowly. It remains analytically distinct because the managed state is the progress of an external task or deployment workflow rather than resource assignment, configuration state, or recovery status. This includes workflow dependencies, execution order, and recovery position. Intent-based multi-cloud storage management~\cite{zheng_intent-based_2025} illustrates this distinction by translating user intent into a structured deployment workflow whose execution and recovery are managed across dependent storage operations.

\subsection{System Contexts}
The preceding decision families do not operate in a neutral environment: the same orchestration problem can change substantially depending on where state is observed, where decisions are interpreted, and where actions take effect. In highly distributed edge--cloud settings, telemetry location, decision placement, network partitioning, and administrative boundaries often force systems to decouple state aggregation, policy validation, and proxy enforcement across different physical or logical nodes~\cite{ding_llm-driven_2026,zheng_intent-based_2025,shastri_llm-driven_2025}.

System context determines how much of the operating environment becomes part of the orchestration decision. We compare it through four linked questions: what is observed, which observations enter the decision context, where the resulting action can take effect, and whether its outcome is fed back into subsequent decisions. Continuum-oriented systems can span several of these boundaries. IntentContinuum, for example, combines application performance with compute and network state and can respond through both resource and network adaptations~\cite{akbari_intentcontinuum_2025}. As supporting evaluation infrastructure, AIOpsLab similarly exposes heterogeneous operational observations together with state-changing interfaces through a common agent–cloud boundary~\cite{shetty_building_2024}. The relevant distinction is therefore not how many edge, cloud, or network components appear in the architecture, but how much of their state and control surface actually participates in the decision.

Physical and temporal conditions further differentiate these contexts. Edge-native formulations incorporate device capacity, communication conditions, or energy state into the feasible action space; in drone offloading~\cite{zhang_ai_2025}, such conditions affect the choice among local execution, edge offloading, and deferred execution. Runtime systems also differ in whether context is used once or remains connected to subsequent control decisions. ARM~\cite{avgerinos_arm_2026} couples remediation with post-action observation and reassessment, allowing updated system state to influence continued intervention. 

\subsection{Connections to Classical Optimization Problems}

The orchestration problems identified in RQ1 expose reusable structures from established optimization and planning problems, but the connection is more useful as a source of solution patterns than as a one-to-one reformulation. Table~\ref{tab:or-connection} therefore links each orchestration task to a classical analog and to LM-assisted approaches that may be transferred from adjacent optimization research. These analogs suggest three transferable patterns: using LMs to translate natural-language specifications into optimization formulations, searching for candidate heuristics under external objectives, and generating plans for checking by model-based verifiers. Such approaches would retain established optimization machinery while using LMs to construct or adapt the representations on which solvers and verifiers operate. 

Transferring these patterns to edge--cloud orchestration introduces additional requirements. Problem formulations must evolve with heterogeneous resources, changing network conditions, distributed observations, service dependencies, and execution feedback. The resulting research opportunity is therefore to couple semantic LM reasoning with optimization components whose feasibility criteria remain explicit and whose models can be revised as operational state changes. Service placement, scheduling, configuration, remediation, and workflow control provide different instances of this broader generate-formalize-verify-adapt pattern.

\input{table/table-op-mapping}

\subsection{Lessons Learned}
Three principal patterns characterize the decision and context landscape. First, LM-based orchestration is predominantly hybrid in formulation. Across resource scheduling, configuration, and resilience, LMs are concentrated in semantic formulation, heuristic or candidate generation, and mitigation planning, while hard constraints, validation, enforcement, and execution commonly remain with downstream system components. This places LMs primarily at the interface between heterogeneous or unstructured information and structured operational mechanisms.

Second, continuum awareness depends on decision constraints, not deployment topology. LLMSched~\cite{ding_llm-driven_2026} uses cluster utilization, task dependencies, and service requirements to guide allocation, whereas drone offloading~\cite{zhang_ai_2025} considers device and communication conditions when choosing local execution, edge offloading, or deferral. IntentContinuum~\cite{akbari_intentcontinuum_2025} further couples application, compute, and network state to resource or network adaptation. These examples distinguish deployment context from continuum conditions that shape feasible actions and subsequent state transitions.

Finally, system context is reported unevenly. Studies commonly identify managed resources and deployment topology, but often omit the observation, timing, actuation, and feedback conditions needed to distinguish episodic artifact generation from environment-aware runtime adaptation or sustained closed-loop control.

%% file: table/table-decision-landscape.tex
\begin{table*}[t]
\centering
\caption{Orchestration decision landscape identified in RQ1.}
\label{tab:decision-landscape}

\scriptsize
\setlength{\tabcolsep}{4pt}
\renewcommand{\arraystretch}{1.08}

\begin{tabularx}{\textwidth}{
    @{}
    >{\raggedright\arraybackslash}p{0.145\textwidth}
    >{\raggedright\arraybackslash}p{0.245\textwidth}
    >{\raggedright\arraybackslash}p{0.265\textwidth}
    >{\raggedright\arraybackslash}X
    @{}
}
\toprule
\textbf{Decision domain}
&
\textbf{Managed decision}
&
\textbf{Typical LM contribution}
&
\textbf{Representative systems}
\\
\midrule

Resource and workload
&
Placement, allocation, routing, offloading, or scheduling of tasks,
services, workloads, and flows under resource and service constraints.
&
Interpret placement preferences; generate candidate assignments or
optimization rewards; in some systems, directly propose an allocation.
&
Semantic Soft Affinity~\cite{sliwko_cluster_2026}; LLM-assisted RL~\cite{tang_llm-assisted_2025}; LLMSched~\cite{ding_llm-driven_2026}; L2D2~\cite{lv_large_2026};
DIRA-LDT~\cite{garcia_llm-based_2025}; adaptive drone offloading~\cite{zhang_ai_2025}.
\\

\addlinespace[2pt]

Configuration and policy
&
Generation or modification of deployment configurations, task
parameters, access-control policies, and security artifacts consumed
by downstream management or enforcement components.
&
Translate intent into machine-readable artifacts; generate or refine
configurations and policies; incorporate validation feedback.
&
Kubernetes manifest generation~\cite{kratzke_dont_2024}; intent-based Kubernetes configuration~\cite{sacco_intent-based_2025}; Web FL~\cite{mawela_web-based_2025}; AgentHAB~\cite{reginold_agenthab_2026}; CollabIoT~\cite{shastri_llm-driven_2025};
LLM2policy~\cite{chen_end--end_2025}; cloud-security orchestration
~\cite{aabirrouche_towards_2026}.
\\

\addlinespace[2pt]

Resilience and operations
&
Diagnosis of degraded states, formulation or execution of corrective
actions, and assessment of the resulting recovery state.
&
Synthesize diagnostic evidence; generate mitigation plans or repair
artifacts; participate in bounded remediation and recovery loops.
&
AIOpsLab~\cite{shetty_building_2024}; AIM~\cite{sarda_leveraging_2026}; low-parameter Kubernetes self-healing~\cite{wiesinger_leveraging_2026};
E2E-REME~\cite{zhang_e2e-reme_2026}; ARM~\cite{avgerinos_arm_2026}; LINEADAPTER~\cite{liu_lineadapter_2025}; LLM-Bayesian RCA ~\cite{pedroso_anomaly_2025}.
\\

\addlinespace[2pt]

Workflow orchestration
&
Ordering and coordination of external deployment or service
operations, including dependencies, execution progress, and recovery
positions.
&
Decompose intent into dependent operations and formulate workflow
structure, while downstream mechanisms control execution and recovery.
&
IBSM
\cite{zheng_intent-based_2025}.
\\

\bottomrule
\end{tabularx}
\vspace{-0.17in}
\end{table*}

%% file: table/table-op-mapping.tex
% \begin{table*}[t]
% \caption{Connections between orchestration tasks and classical optimization problems.}
% \label{tab:or-connection}
% \centering
% \footnotesize
% \setlength{\tabcolsep}{3.5pt}
% \renewcommand{\arraystretch}{1.05}
% \begin{tabularx}{\textwidth}{
%     >{\raggedright\arraybackslash}p{0.17\textwidth}
%     >{\raggedright\arraybackslash}p{0.22\textwidth}
%     >{\raggedright\arraybackslash}p{0.22\textwidth}
%     >{\raggedright\arraybackslash}X}
% \toprule
% \textbf{Task} &
% \textbf{Optimization problem} &
% \textbf{Possible LM role} &
% \textbf{Edge--cloud considerations} \\
% \midrule

% Service placement &
% Facility location; generalized assignment &
% Intent and constraint formulation &
% Capacity, latency, dependencies, failure domains \\

% Workload allocation &
% Multidimensional bin packing; online allocation &
% Preference/constraint formulation; candidate generation &
% Heterogeneity, dynamic arrivals, SLOs, migration cost \\

% Configuration / policy &
% Constraint satisfaction &
% Constraint or configuration generation &
% Cross-resource constraints, permissions, runtime state \\

% Remediation &
% Sequential planning &
% Candidate action or plan generation &
% Partial observability, side effects, recovery failure \\

% Workflow orchestration &
% Precedence-constrained scheduling &
% Dependency or plan formulation &
% Cross-site dependencies, partial failure, checkpoints \\

% \bottomrule
% \end{tabularx}
% \end{table*}

\begin{table*}[t]
\caption{Classical problem analogues and transferable LM--optimization
opportunities for edge--cloud orchestration.}
\label{tab:or-connection}
\centering
\scriptsize
\setlength{\tabcolsep}{3.5pt}
\renewcommand{\arraystretch}{1.06}

\begin{tabularx}{\textwidth}{
    >{\raggedright\arraybackslash}p{0.13\textwidth}
    >{\raggedright\arraybackslash}p{0.17\textwidth}
    >{\raggedright\arraybackslash}p{0.30\textwidth}
    >{\raggedright\arraybackslash}X}
\toprule
\textbf{Orchestration task} &
\textbf{Classical analogue} &
\textbf{Transferable LM--optimization pattern} &
\textbf{Edge--cloud research opportunity} \\
\midrule

Service placement &
Capacitated facility location;
generalized assignment
&
Translate semantic objectives and hard/soft requirements into
solver-ready variables, constraints, and objectives.
&
Continuously reformulate placement as capacity, latency,
service dependencies, and failure-domain conditions change,
while externally checking feasibility against live topology.
\\

Workload scheduling / offloading &
Multidimensional bin packing;
online scheduling
&
Generate or evolve scheduling heuristics, scoring rules, or
candidate actions and evaluate them with an external solver,
simulator, or objective function.
&
Adapt scheduling logic to workload shifts, heterogeneous
resources, SLOs, migration cost, and energy state without
placing an LM on every latency-critical decision path.
\\

Configuration / policy &
Constraint satisfaction;
constraint programming
&
Generate formal constraints or candidate configurations and use
an external verifier to reject conflicts and provide feedback for
iterative repair.
&
Move from static syntax checking toward state-aware
generate--verify--repair loops over live resources, permissions,
dependencies, and policy interactions.
\\

Remediation &
Constrained sequential planning;
stochastic shortest path
&
Generate candidate recovery plans while model-based checks,
simulation, or runtime feedback assess preconditions,
side effects, and goal satisfaction.
&
Plan under partial observability and action risk while accounting
for recovery cost, failed interventions, rollback, fallback, and
changing system state.
\\

Workflow orchestration &
Precedence-constrained scheduling;
resource-constrained project scheduling
&
Derive task graphs, precedence relations, and candidate plans
from high-level intent, while a scheduler or verifier retains
control over feasible execution.
&
Support dynamic replanning across sites after partial failures
while preserving dependencies, resource constraints,
checkpoints, and recovery state.
\\

\bottomrule
\end{tabularx} 
\vspace{-0.17in}
\end{table*}

%% file: sections/600-rq2.tex
RQ2 examines how LM contributions are translated into operational control. We first distinguish the LM's decision responsibility, then trace the handoffs through which its outputs pass downstream. We then examine which consequential rights the LM is granted and which remain with optimizers, validators, executors, policy mechanisms, or operators, before analyzing safeguards that constrain state-changing actions. Table~\ref{tab:responsibility-chain} maps these distinctions onto the orchestration decision chain.

\subsection{LM Roles}\label{LMRolee}

\begin{figure*}
    \centering
    \includegraphics[width=0.8\linewidth]{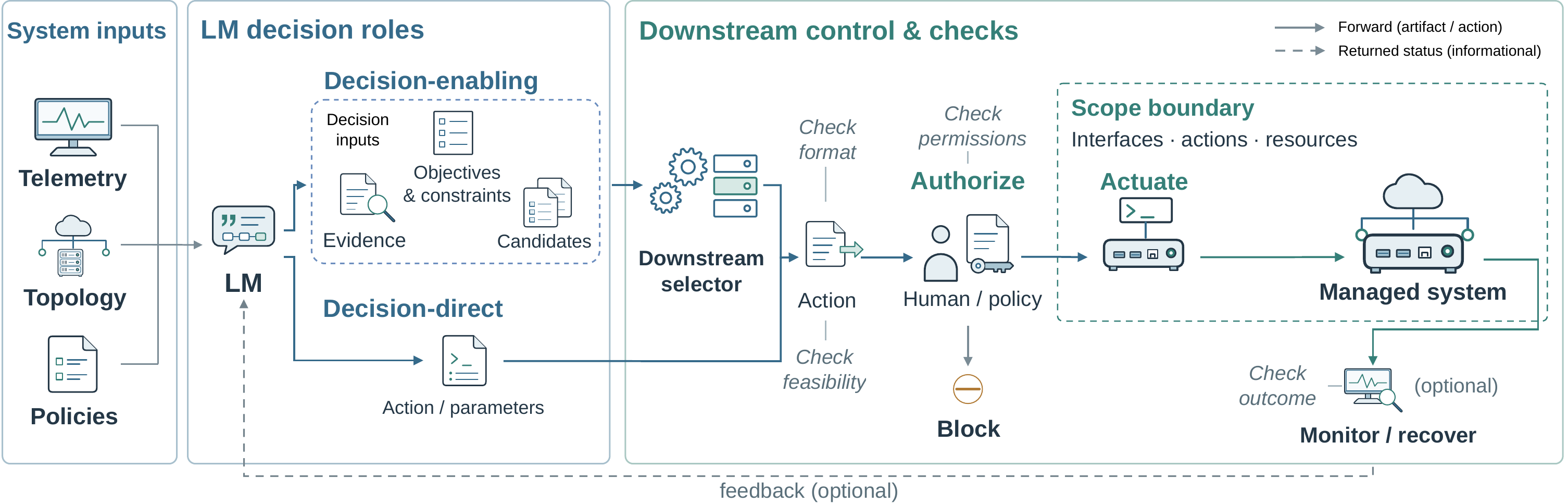}
    \caption{LM decision roles and handoffs to operational control.}
    \label{fig:rq2}
    \vspace{-0.17in}
\end{figure*}

\input{table/table-responsibility}

LM participation differs first in whether the model shapes a downstream decision or specifies the action itself. We distinguish decision-enabling roles, in which the LM structures evidence, formulates objectives, or generates candidates for downstream selection, from decision-direct roles, in which the LM specifies a proposed action, plan, policy, or key operational parameter. Either role may be one-shot or closed-loop: post-action observations can be incorporated into subsequent diagnosis, planning, or remediation. Role therefore describes decision responsibility rather than the authority to select, authorize, or execute the resulting action.

Decision-enabling roles appear in optimization-oriented systems~\cite{ding_llm-driven_2026,tang_llm-assisted_2025,lv_large_2026} and diagnostic workflows. LMs may formulate causal hypotheses~\cite{tang_root_2025}, interpret narrowed evidence~\cite{li_magmascope_2026,pu_errorprism_2025}, or reason over profiling data~\cite{ye_profrca_2026}, code~\cite{li_coca_nodate}, classifier outputs~\cite{han_potential_2024}, and multimodal observations~\cite{zhang_tamofine-grained_2025}. 
Monitoring-oriented systems can remain upstream: LEMON~\cite{gotz_lemon_2026} uses LMs to interpret monitoring intents and synthesize telemetry-backed responses, while the evaluated workflow excludes state-changing actuation. Even extensive internal autonomy, including hierarchical planning~\cite{gao_rcaflow_nodate}, recursive reasoning, memory~\cite{zhang_agentic_2026}, or multi-agent coordination~\cite{gao_rcaflow_nodate,parthasarathy_engineering_2025}, may still terminate in information or recommendations rather than infrastructure change.

Decision-direct roles occur when the LM specifies corrective or reconfiguration actions, as in intent-driven resource adaptation~\cite{akbari_intentcontinuum_2025}, bounded infrastructure remediation~\cite{avgerinos_arm_2026}, or generated recovery plans~\cite{zhang_e2e-reme_2026}. Some systems keep the LM in the feedback loop, using execution outcomes or updated system state to revise subsequent remediation~\cite{avgerinos_arm_2026,wiesinger_leveraging_2026,zhang_e2e-reme_2026}. This extends decision participation across successive state changes, not authority by itself.

Implementation choices do not map directly onto these roles. The corpus includes hosted foundation models~\cite{sliwko_cluster_2026,sarda_leveraging_2026}, fine-tuned compact models~\cite{sacco_intent-based_2025,gotz_lemon_2026,liu_lineadapter_2025}, and locally deployed low-parameter models~\cite{wiesinger_leveraging_2026}, with adaptation ranging from prompting~\cite{kratzke_dont_2024} and retrieval~\cite{reginold_agenthab_2026} to task-specific fine-tuning~\cite{liu_lineadapter_2025}. These choices affect latency, privacy, cost, and deployability more directly than operational authority: a local model~\cite{wiesinger_leveraging_2026} may participate in remediation, while a larger hosted model~\cite{sarda_leveraging_2026} may remain advisory.

\subsection{Handoffs}
Handoffs identify where LM outputs pass to downstream components without necessarily transferring decision authority. We classify them by two features: what is transferred and which downstream decision or execution functions remain to be performed. Decision-shaping handoffs transfer objectives, evidence, or candidates; control-artifact handoffs transfer an operational configuration or policy; and execution handoffs transfer an intended action to a privileged executor. These patterns locate the control boundary rather than define a monotonic autonomy scale.

Decision-shaping handoffs leave the consequential choice downstream. LLMSched transfers candidate schedules to an optimization stage~\cite{ding_llm-driven_2026}, lightweight LLM-assisted RL supplies candidate actions to a DQN~\cite{tang_llm-assisted_2025}, and L2D2~\cite{lv_large_2026} provides an LM-designed reward function for subsequent reinforcement learning. In each case, the LM constrains or reshapes the decision space, while a non-LM mechanism retains final selection or feasibility enforcement.

Control-artifact handoffs move closer to operation: LM output specifies a configuration or policy, but downstream components still determine how it enters the managed system. Generated OpenHAB~\cite{reginold_agenthab_2026} rules pass to validation, CollabIoT~\cite{shastri_llm-driven_2025} policies to configuration and enforcement mechanisms, LLM2policy~\cite{chen_end--end_2025} artifacts to service-mesh policy control, Sigma rules to SIEM~\cite{aabirrouche_towards_2026} processing, and federated-learning configurations to the parameter server~\cite{mawela_web-based_2025}. In these cases, the recipient determines whether the artifact is validated, adopted, translated, or enforced.

Execution handoffs move closest to state change: the LM specifies the intended action, while a separate orchestrator, tool interface, or executor performs privileged actuation~\cite{akbari_intentcontinuum_2025,avgerinos_arm_2026,wiesinger_leveraging_2026,zhang_e2e-reme_2026}. Handoff structure reveals how much consequential
responsibility remains downstream and links LM roles to operational authority. 

\subsection{Authority Profiles}

Authority concerns which consequential rights the LM actually has. We distinguish action selection, authorization, privileged actuation, and post-action recovery, while treating tool, resource, and action scope as boundaries on those rights. An LM may therefore produce an operationally complete schedule, policy, or remediation plan while final selection, approval, execution, or recovery remains elsewhere. 

At the advisory end, the LM influences a decision without controlling the resulting state transition. Semantic soft-affinity parsing~\cite{sliwko_cluster_2026} contributes structured preferences to scheduler scoring, while LLMSched~\cite{ding_llm-driven_2026} delegates feasibility enforcement and constrained selection to an optimization stage.

Between advisory influence and bounded execution, a second pattern allows LM-generated or LM-selected actions to progress toward execution while retaining a separate control boundary. AIOpsLab~\cite{shetty_building_2024} places an orchestrator between agent requests and privileged infrastructure operations, and low-parameter Kubernetes self-healing~\cite{wiesinger_leveraging_2026} separates LM-generated remediation from the component that applies it. AIM~\cite{sarda_leveraging_2026} extends mitigation planning to executable playbooks in a controlled testbed, while retaining a human-in-the-loop operational model. 

At the higher-authority end, execution is implemented through explicit operational bounds. ARM~\cite{avgerinos_arm_2026} permits the agent to select and invoke infrastructure actions through a scoped management interface, while E2E-REME~\cite{zhang_e2e-reme_2026} links remediation generation to execution and verification within a defined recovery workflow. In both cases, state-changing responsibility is exercised within defined action, resource, and execution scopes.

Authority can also vary by action risk. TAR~\cite{hu_novel_2025} automates lower-risk responses through SOAR and cloud APIs while routing higher-risk threats to security experts. A Dual-LM drone architecture~\cite{zhang_ai_2025} similarly separates initial action generation from risk-constrained finalization. Authority is therefore not captured by labels such as ``agentic'' or ``autonomous''; the decisive distinction is which consequential rights are actually transferred, under what operational boundaries, and for which classes of actions.

\subsection{Operational Assurance}\label{OA}

Operational assurance concerns both what property is checked and how the check constrains progression toward state change. We distinguish four assurance targets: representation validity, decision validity, execution safety, and outcome/recovery validity.

Representation assurance is most visible in configuration-oriented systems. Generated Kubernetes manifests are evaluated for deployability~\cite{kratzke_dont_2024}, while AgentHAB~\cite{reginold_agenthab_2026} uses parser-based syntax checking within an iterative generation loop. CollabIoT~\cite{shastri_llm-driven_2025} combines structured-output validation with semantic checking and a programmatic attribute validator that queries the current policy and device state.

Decision assurance becomes more consequential when LM outputs participate directly in allocation or action selection. LLMSched passes LM-generated scheduling candidates to an ILP-based refinement stage~\cite{ding_llm-driven_2026}, while LLM-assisted RL supplies candidate actions to a downstream DQN~\cite{tang_llm-assisted_2025}. Semantic Soft Affinity integrates interpreted preferences with scheduler-side scoring~\cite{sliwko_cluster_2026}, and DIRA-LDT~\cite{garcia_llm-based_2025} evaluates recommended allocations against resource and latency constraints before execution. Across these designs, assurance depends on whether the downstream mechanism merely influences selection or enforces an explicit feasibility condition that can reject an LM-proposed choice. 

Execution assurance concerns the transition from an accepted action to privileged system change. ARM~\cite{avgerinos_arm_2026} exposes infrastructure operations through scoped tool interfaces, with safeguards and human approval configurable per action. TAR~\cite{hu_novel_2025} makes the boundary risk-dependent: lower-risk responses can proceed through SOAR and cloud APIs, whereas higher-risk cases are retained for expert investigation. Aabirrouche et al.~\cite{aabirrouche_towards_2026} provide a complementary pre-deployment pattern in which generated security artifacts are validated before activation. Together, these systems illustrate interface restriction, action-specific approval, and pre-execution validation as distinct forms of execution assurance.

Outcome and recovery assurance extends control beyond actuation. ARM~\cite{avgerinos_arm_2026} evaluates post-intervention metrics before concluding recovery, IntentContinuum~\cite{akbari_intentcontinuum_2025} continues monitoring whether the target intent is restored, and IBSM~\cite{zheng_intent-based_2025} retains execution checkpoints for workflow recovery. E2E-REME~\cite{zhang_e2e-reme_2026} similarly couples execution with verification and refinement. Post-action monitoring confirms whether the intended state is reached, whereas check-pointing or iterative refinement supports continued recovery when it is not. These mechanisms distinguish successful actuation from successful recovery: an action may execute correctly without restoring the intended operating condition. 

Evaluator independence and blocking power cut across these four assurance targets. Self-refinement or verification by the same or another LM may improve consistency, but it is not equivalent to a programmatic validator, solver, policy mechanism, runtime oracle, or human reviewer that establishes an external acceptance condition. The operational significance of a check further depends on whether it can prevent, redirect, or reverse the corresponding action.

\subsection{Lessons Learned}

Three findings emerge from the authority analysis. First, LM responsibility generally extends further than LM authority: models interpret evidence, formulate objectives, generate candidates, and in some cases specify actions, while final selection, authorization, or privileged execution often remains downstream. Second, this separation is enforced through different control boundaries—optimizers and schedulers in resource management, validators and policy mechanisms in configuration, and tool interfaces or human gates in remediation~\cite{sliwko_cluster_2026, ding_llm-driven_2026, shastri_llm-driven_2025, avgerinos_arm_2026, hu_novel_2025}. Third, operational authority is granular, not agent-wide: it depends on action scope, risk, and permissions for consequential operations. Agent count or planning depth is therefore not a proxy for infrastructure authority. Across the reviewed studies, the highest-authority profile supported by the evidence is bounded control within explicit tool, action, and recovery scopes. 

%% file: table/table-responsibility.tex
\begin{table*}[t]
\centering
\caption{LM responsibility and downstream control across the orchestration decision chain.}
\label{tab:responsibility-chain}

\scriptsize
\setlength{\tabcolsep}{3.8pt}
\renewcommand{\arraystretch}{1.06}

\begin{tabularx}{\textwidth}{
    @{}
    >{\raggedright\arraybackslash}p{0.13\textwidth}
    >{\raggedright\arraybackslash}p{0.28\textwidth}
    >{\raggedright\arraybackslash}p{0.33\textwidth}
    >{\raggedright\arraybackslash}X
    @{}
}
\toprule
\textbf{Stage}
&
\textbf{Functional contribution}
&
\textbf{Downstream control}
&
\textbf{Representative systems}
\\
\midrule

Observe/Ground
&
Acquire or assemble decision-relevant intent, telemetry, system state,
history, or retrieved evidence.
&
Grounded state becomes input to interpretation, diagnosis, or planning;
observation itself does not determine an action.
&
LEMON~\cite{gotz_lemon_2026}; IntentContinuum~\cite{akbari_intentcontinuum_2025}; E2E-REME~\cite{zhang_e2e-reme_2026}; ARM~\cite{avgerinos_arm_2026}.
\\

\addlinespace[1.5pt]

Interpret/Formulate
&
Translate intent or evidence into a decision representation, objective,
preference, diagnosis, or structured task.
&
A downstream planner, optimizer, or executor still determines how the
representation affects the managed state.
&
Semantic Soft Affinity~\cite{sliwko_cluster_2026}; Web FL~\cite{mawela_web-based_2025}.
\\

\addlinespace[1.5pt]

Generate/Plan
&
Produce candidate actions, schedules, configurations, policies, rewards,
or remediation plans.
&
Candidates may still require validation, selection, authorization, or
execution by another component.
&
LLMSched~\cite{ding_llm-driven_2026}; LLM-assisted RL~\cite{tang_llm-assisted_2025}; AgentHAB~\cite{reginold_agenthab_2026}; AIM~\cite{sarda_leveraging_2026}.
\\

\addlinespace[1.5pt]

Validate
&
Check structural, semantic, constraint, policy, or execution properties
and return acceptance or refinement feedback.
&
Validation may block or redirect an output but does not itself imply
authorization to execute it.
&
AgentHAB~\cite{reginold_agenthab_2026}; LLMSched~\cite{ding_llm-driven_2026}; AIOpsLab~\cite{shetty_building_2024}; CollabIoT~\cite{shastri_llm-driven_2025}.
\\

\addlinespace[1.5pt]

Select
&
Choose an action or artifact from generated or feasible alternatives.
&
The selected option may still require policy or human authorization
before actuation.
&
LLM-assisted RL~\cite{tang_llm-assisted_2025}; LLMSched~\cite{ding_llm-driven_2026}; DIRA-LDT~\cite{garcia_llm-based_2025}; adaptive drone offloading~\cite{zhang_ai_2025}.
\\

\addlinespace[1.5pt]

Authorize
&
Permit a selected action within an approved resource, tool, or risk scope.
&
Authorization defines whether and under what scope an executor may
perform the state change.
&
ARM~\cite{avgerinos_arm_2026}; TAR~\cite{hu_novel_2025}.
\\

\addlinespace[1.5pt]

Act
&
Apply an authorized configuration, policy, playbook, or management
operation to the target system.
&
Privileged actuation may remain with an orchestrator, tool interface,
policy plane, or other executor rather than the LM itself.
&
AIOpsLab~\cite{shetty_building_2024}; ARM~\cite{avgerinos_arm_2026}; E2E-REME~\cite{zhang_e2e-reme_2026}; CollabIoT~\cite{shastri_llm-driven_2025}.
\\

\addlinespace[1.5pt]

Monitor/Recover
&
Assess post-action state and support retry, replanning, fallback,
rollback, checkpoint recovery, or escalation.
&
Observed outcomes determine whether the loop terminates, revises the
action, or transfers control to another recovery mechanism.
&
E2E-REME~\cite{zhang_e2e-reme_2026}; ARM~\cite{avgerinos_arm_2026}; IntentContinuum~\cite{akbari_intentcontinuum_2025}; IBSM~\cite{zheng_intent-based_2025}.
\\

\bottomrule
\end{tabularx} 
\vspace{-0.2in}
\end{table*}

%% file: sections/700-rq3.tex
\begin{figure*}
    \centering
    \includegraphics[width=0.8\linewidth]{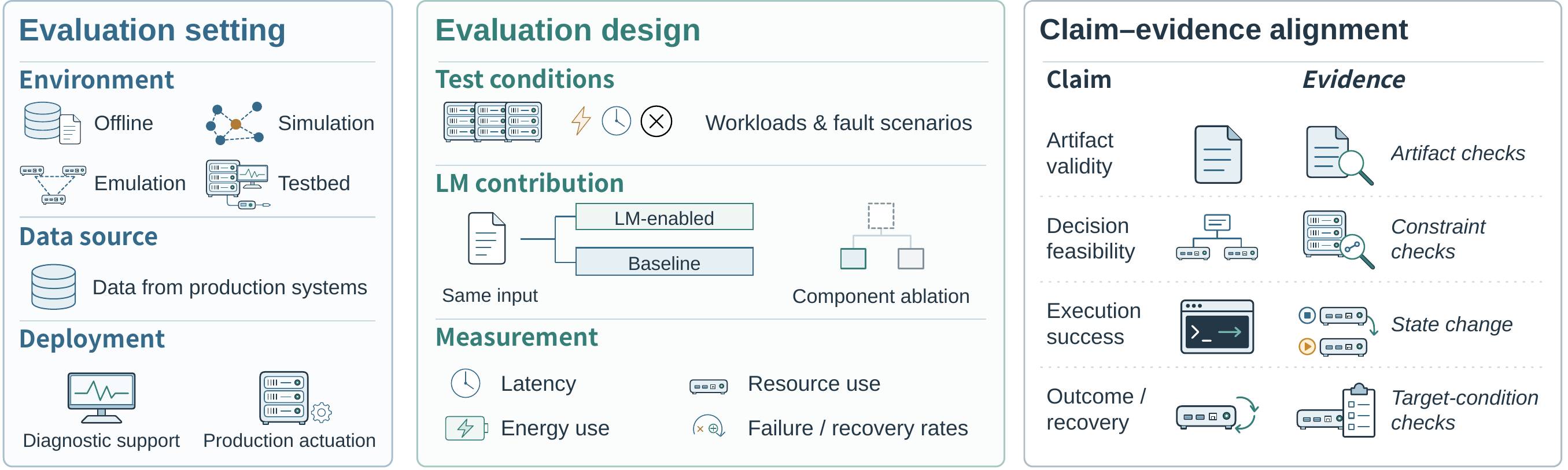}
    \caption{Evaluation and deployment settings, study design, and reported orchestration outcomes.}
    \label{fig:rq3}
    \vspace{-0.15in}
\end{figure*}

RQ3 examines what orchestration effects the reported evaluations establish, under which operating conditions, and with what attribution to the LM-enabled design. Figure~\ref{fig:rq3} organizes this assessment into evaluation setting, evaluation design, and claim-evidence alignment. We then examine security and sustainability evidence as cross-cutting concerns.

\subsection{Evidence Maturity}

Evidence maturity has two independent dimensions: the evaluation environment (offline or simulation, controlled execution, or operational deployment) and the endpoint reached, from artifact or decision quality to state change and post-action outcome. Production-derived data strengthen diagnostic provenance, whereas actuation evidence requires evaluated state changes. Workload scale, duration, dynamics, and fault provenance further bound the claims. 

Many decision-oriented studies provide strong task-local evidence in simulation or bounded experimental settings. L2D2~\cite{lv_large_2026} reports improved energy efficiency for multi-UAV MEC scheduling, LLMSched~\cite{ding_llm-driven_2026} evaluates scheduling gains on production-derived cluster traces, and Web FL~\cite{mawela_web-based_2025} measures communication and computational improvements for automated federated-learning tasks. These experiments establish local decision quality or efficiency within the evaluated formulation, with claim scope defined by the tested topology, workload distribution, action space, and execution duration.

Controlled execution environments extend evidence to concrete system interfaces and state changes. CollabIoT~\cite{shastri_llm-driven_2025} evaluates policy-driven device configuration on an IoT testbed, Semantic Soft Affinity~\cite{sliwko_cluster_2026} integrates LM-derived preferences with a Kubernetes/OpenShift scheduling path, and ARM~\cite{avgerinos_arm_2026} executes remediation actions on a controlled cloud–edge K3S environment. Such studies provide stronger evidence for operational integration because generated or selected actions reach the managed system under explicit resource, workload, fault, and action vocabularies.

At the operational end, evidence is concentrated in diagnostic and decision-support systems. RCACopilot~\cite{chen_automatic_2024} evaluates RCA on Microsoft incident data and reports long-term production use of its diagnostic information collection component. ErrorPrism~\cite{pu_errorprism_2025} reports production deployment for error-path reconstruction, while AlertGuardian~\cite{yu_alertguardian_2025} reports deployment or pilot use at different stages of its alert-management pipeline. Marcano et al.~\cite{marcano_scaling_2026} provide a different form of operational evidence through production-scale resilience validation with fault injection restricted to isolated test traffic. These studies support operational diagnosis, decision support, or resilience validation without necessarily extending the evaluated endpoint to autonomous infrastructure actuation. 

\subsection{Claim-Evidence Alignment}\label{CEA}

\begin{figure}
    \centering
    \includegraphics[width=\linewidth]{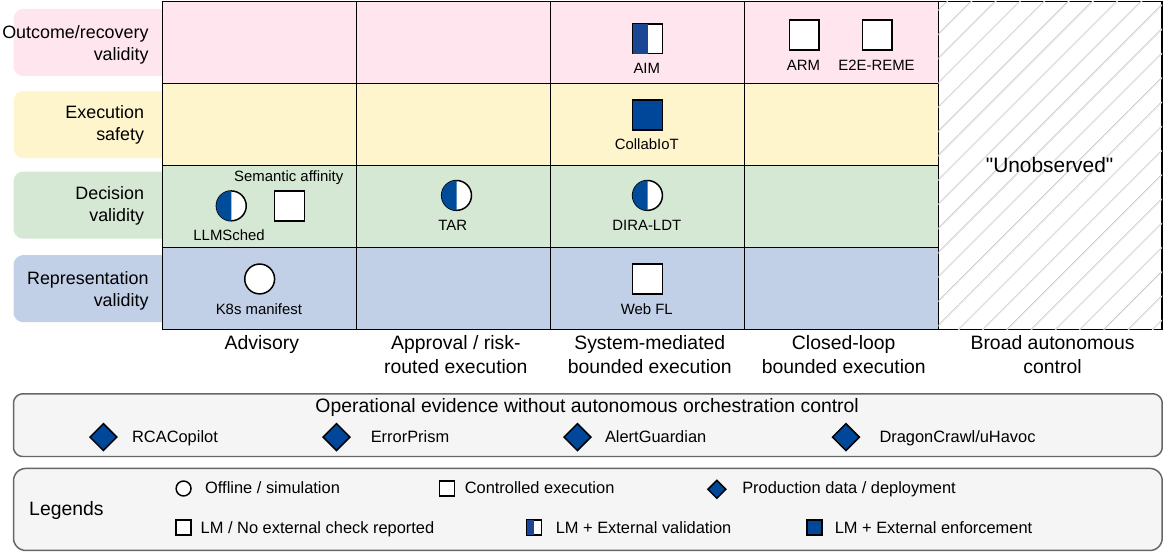}
    \caption{Matrix of authority, assurance, and evidence in LM-enabled orchestration.}
    \label{fig:authority-assurance-evidence}
    \vspace{-0.15in}
\end{figure}

Claim--evidence alignment depends on the evaluated endpoint. Figure~\ref{fig:authority-assurance-evidence} illustrates representative control profiles and assessed properties, not mutually exclusive system classes or reliability guarantees: approval gates, execution mediation, and feedback may coexist.

Configuration studies illustrate how evidence can progress beyond artifact validity. Kubernetes manifest generation~\cite{kratzke_dont_2024} primarily evaluates whether LM-produced artifacts are correct and deployable. LLM2policy~\cite{chen_end--end_2025} extends the endpoint with policy-level tests and attack simulation, while CollabIoT~\cite{shastri_llm-driven_2025} evaluates generated policies together with runtime configuration and enforcement on an IoT testbed. The supported claim therefore depends less on the artifact type itself than on how far the evaluation follows that artifact into the managed system.

Scheduling studies extend the endpoint into the decision path. Semantic Soft Affinity~\cite{sliwko_cluster_2026} evaluates both interpretation accuracy and resulting placement behavior through an integrated scheduler prototype. LLMSched~\cite{ding_llm-driven_2026} evaluates LM-generated candidates after downstream constraint enforcement and further uses component ablations to isolate contributions within its hybrid LM--optimization design. Together, these designs substantiate bounded allocation decisions under the evaluated workloads and constraint models.

Closed-loop remediation requires post-action evidence covering failed, partial, and successful outcomes. ARM~\cite{avgerinos_arm_2026} evaluates whether selected interventions restore service conditions, while E2E-REME~\cite{zhang_e2e-reme_2026} executes generated playbooks and verifies post-repair state. AIM~\cite{sarda_leveraging_2026} explicitly distinguishes task completion from recovery: generated playbooks achieve a 71.9\% task-completion rate, whereas binary remediation success is 25\%. Together, these results show that action completion and system recovery are distinct evaluation endpoints.

Evidence strength also depends on what the comparison isolates and where the cost boundary is drawn. In hybrid systems, end-to-end gains may arise from the LM together with retrieval, optimization, validation, or execution components; component ablations, such as those reported for LLMSched, provide stronger attribution than system-level baselines alone~\cite{ding_llm-driven_2026}. Cost measurements likewise cover different portions of the control path: CollabIoT~\cite{shastri_llm-driven_2025} reports configuration and enforcement overheads, LLM2policy~\cite{chen_end--end_2025} measures policy-generation and update latency, and LEMON~\cite{gotz_lemon_2026} decomposes response latency across individual pipeline stages and the end-to-end workflow. Operational deployment therefore does not remove the need to distinguish diagnostic support from actuation evidence~\cite{chen_automatic_2024,pu_errorprism_2025,yu_alertguardian_2025}. Claim--evidence alignment therefore depends jointly on endpoint reach, attribution, failure and recovery behavior, and the measured cost boundary.

\subsection{Security Evidence}\label{SE}

Security introduces a further distinction: using LMs for security-oriented orchestration is different from showing that the LM-enabled orchestration process is itself secure. For the former, several studies evaluate security tasks beyond language-level output quality. Infrastructure as Code (IaC) and Kubernetes studies assess vulnerability or misconfiguration detection and remediation quality~\cite{toprani_llm_2025,ghorab_towards_2025}. 

CollabIoT~\cite{shastri_llm-driven_2025} evaluates generated access-control policies together with runtime enforcement and associated overheads, while LLM2policy~\cite{chen_end--end_2025} deploys generated service-mesh policies, tests their structural and operational validity, and uses attack simulations to assess whether unauthorized interactions are blocked. Cloud-security orchestration~\cite{aabirrouche_towards_2026} further validates generated security and detection rules in contained environments, and TAR~\cite{hu_novel_2025} evaluates threat classification and risk-dependent response, automating lower-risk cases while escalating higher-risk threats. These evaluations substantiate security-oriented orchestration within the tested policy classes, threat scenarios, enforcement mechanisms, and execution environments.

A different question is whether the LM-enabled orchestration mechanism is itself evaluated as a security-sensitive control surface. Few evaluations examine this control path directly. Some systems reduce exposure through bounded interfaces, external policy enforcement, or risk-dependent human escalation; LEMON~\cite{gotz_lemon_2026} also explicitly avoids executable query generation partly to reduce indirect prompt-injection risk. The reviewed evaluations predominantly test whether security artifacts or responses are correct and effective; adversarial testing of malicious intents, poisoned telemetry or retrieved context, agent or tool hijacking, privilege escalation through generated actions, and information leakage is rarely reported. Security evidence should therefore distinguish using LMs to secure an orchestrated system from establishing the security of LM-enabled orchestration itself; evidence for the former does not establish the latter.

\subsection{Sustainability Evidence and SDG Alignment}

\input{table/table-sdg-mapping}

The reviewed studies connect most clearly to UN Sustainable Development Goal (SDG) targets on energy efficiency, resilient infrastructure, and resource-use efficiency (Table~\ref{tab:sdg}). Reported evidence includes energy-aware scheduling \cite{lv_large_2026}, verified recovery \cite{avgerinos_arm_2026,zhang_e2e-reme_2026}, and communication or computational savings \cite{mawela_web-based_2025}. These results establish benefits within the evaluated workload or control process. Assessing net sustainability, however, requires accounting for LM inference, validation, retries, and communication alongside those benefits. Carbon intensity and hardware-lifecycle effects are rarely measured. We therefore distinguish alignment with an SDG objective from demonstrated environmental benefit and use the missing measures to identify evaluation priorities.

\subsection{Lessons Learned}
Three findings characterize the evaluation landscape. First, most studies evaluate artifact and decision quality, while a smaller subset verifies post-action recovery. Second, environmental realism and endpoint reach vary independently: production studies substantiate diagnosis and alert management~\cite{chen_automatic_2024,pu_errorprism_2025,yu_alertguardian_2025}, whereas controlled testbeds evaluate bounded actuation or recovery~\cite{avgerinos_arm_2026,shastri_llm-driven_2025,zhang_e2e-reme_2026}. Third, the strongest actuation evidence establishes scoped operational control within tested workload, fault, tool, and topology envelopes. Across RQ1--RQ3, SPACE reveals that assurance is shaped by the managed decision rather than by autonomy level alone. Scheduling candidates undergo constraint-based feasibility checks~\cite{ding_llm-driven_2026}, generated policies pass through validation and enforcement~\cite{shastri_llm-driven_2025,chen_end--end_2025}, and remediation outcomes require post-action recovery checks~\cite{avgerinos_arm_2026,zhang_e2e-reme_2026}. These requirements also depend on continuum conditions: ARM shows that limited edge capacity restricts available remedies and can prevent full baseline restoration~\cite{avgerinos_arm_2026}. Dependable orchestration therefore requires aligning decision constraints and operating context (RQ1), LM responsibility and downstream controls (RQ2), and evidence for the claimed system effect (RQ3).

%% file: table/table-sdg-mapping.tex
% \begin{table*}[t]
% \caption{Indicative alignment of reported orchestration objectives
% with selected UN SDG targets.}
% \label{tab:sdg}
% \centering
% \footnotesize
% \setlength{\tabcolsep}{4pt}
% \renewcommand{\arraystretch}{1.05}
% \begin{tabularx}{\textwidth}{
%     >{\raggedright\arraybackslash}p{0.18\textwidth}
%     >{\raggedright\arraybackslash}p{0.34\textwidth}
%     >{\raggedright\arraybackslash}X}
% \toprule
% \textbf{SDG target} &
% \textbf{Reported evidence} &
% \textbf{Main gap} \\
% \midrule

% 7.3 Energy efficiency &
% Energy-aware scheduling and task allocation &
% Net energy cost including LM and control overhead \\

% 9.1 Resilient infrastructure &
% Fault mitigation, self-healing, recovery verification &
% Evidence across broader faults and operating conditions \\

% 9.4 Resource-use efficiency &
% Resource utilization, communication, and computation efficiency &
% Link system efficiency to energy or environmental outcomes \\

% 12.2 Natural-resource efficiency &
% No direct material or hardware-lifecycle evidence identified &
% Hardware lifetime and material-footprint measures \\

% \bottomrule
% \end{tabularx}
% \end{table*}

% Requires: booktabs, tabularx, array
\begin{table}[t]
\caption{Alignment of reported orchestration objectives
with selected UN SDG targets.}% ~\cite{un2030agenda}.}
\label{tab:sdg}
\centering
\scriptsize
\setlength{\tabcolsep}{3pt}
\renewcommand{\arraystretch}{1.05}
\begin{tabularx}{\columnwidth}{
    @{}
    >{\raggedright\arraybackslash}p{0.27\columnwidth}
    >{\raggedright\arraybackslash}X
    >{\raggedright\arraybackslash}X
    @{}}
\toprule
\textbf{SDG target} &
\textbf{Reported evidence} &
\textbf{Main gap} \\
\midrule

\textbf{7.3} Energy efficiency &
Energy-aware scheduling and allocation &
Net energy savings after LM and control overhead \\

\textbf{9.1} Infrastructure resilience &
Fault mitigation, self-healing, and recovery verification &
Coverage of diverse faults and operating conditions \\

\textbf{9.4} Resource-use efficiency &
Efficient resource use, communication, and computation &
Measured energy or environmental benefits \\

\textbf{12.2} Natural-resource efficiency &
No direct material or hardware-lifecycle evidence identified &
Hardware lifetime and material footprint \\

\bottomrule
\end{tabularx} 
\vspace{-0.17in}
\end{table}

%% file: sections/800-future.tex
Section~\ref{section/rq1} identifies limited incorporation of continuum dynamics into orchestration decisions. The unresolved problem is maintaining feasible placement, offloading, and workflow execution when resource and network observations are delayed or incomplete. The research should thus investigate intent-to-constraint compilation coupled with robust optimization under uncertainty in resource capacity and communication delay. Generated constraints must preserve intent semantics, while independent feasibility checks account for service dependencies, locality, migration overhead, and workflow recovery state. The central question is how observation age and replanning frequency affect constraint violations and SLO satisfaction.

The bounded interpretation and candidate-selection roles identified in Section~\ref{LMRolee} raise a second problem: deciding when generative inference is necessary within the control path. JEV’s lower decision latency and declining accuracy on wider contracts~\cite{li2026replacing} expose a tradeoff between decision speed and joint semantic correctness. NetLLM~\cite{wu2024netllm} demonstrates task-specific output heads for networking decisions. Future architectures could combine such specialization with typed decision models and generative planning for complex requests. Extending model routing~\cite{ong2025routellm} requires calibrated estimates of joint decision error, action risk, queue occupancy, and remaining deadline slack. Research must thus establish which decision stages can be specialized while preserving correct, on-time service completion.

Sections~\ref{OA} and~\ref{SE} expose an assurance gap between validating an output and controlling its effects. A checked action can become infeasible before execution, while malicious operational context can induce a well-formed but unauthorized request. The research should thus investigate action contracts with explicit preconditions, permission scopes, and postconditions, bound to state versions or resource reservations. Independent runtime monitors should verify effects and initiate compensation or escalation after partial execution. Evaluation must test concurrent controller interference, network partitions, and prompt injection through their resulting unauthorized state changes and recovery failures.

Future research should also investigate parameter-efficient adaptation approaches like QLoRA~\cite{dettmers2023qlora} and execution-grounded reinforcement fine-tuning~\cite{zhang_e2e-reme_2026} for orchestration-specific models. Training should use verified state-action-outcome trajectories, with rewards reflecting sustained SLO satisfaction, successful recovery, and intervention cost. Key challenges thus include credit assignment under delayed feedback and partial execution, and generalization across unseen fault combinations, topologies, and connectivity conditions. Evaluation should measure cumulative SLO violations, recovery persistence, and total orchestration cost and energy, including inference, validation, retries, communication, and amortized model adaptation. Component ablations should isolate the contributions of the model, verifier, and execution controller.

%% file: sections/900-conclusion.tex
This survey examined LM-enabled service orchestration across the edge--cloud continuum using the orchestration decision as the unit of analysis, linking managed state and continuum constraints to LM responsibility, operational authority, assurance, and empirical evidence. This perspective distinguishes deployment context from decision-relevant continuum conditions, action specification from authority to execute, and successful actuation from verified recovery. The reviewed literature reveals a predominantly hybrid design space in which LMs interpret intents, generate candidates, or propose corrective actions, while downstream components commonly retain feasibility enforcement and privileged control. Although bounded closed-loop remediation systems are emerging, production-scale evidence of closed-loop control across heterogeneous edge--cloud environments remains limited. These findings suggest that reliance on LM-enabled orchestration depends on aligning decision formulation, operational authority, safeguards, and evaluation endpoints with the claimed system effect, rather than on agent autonomy alone.

%% file: sections/1099-supp.tex
\section{Search and Screening Details}
\label{app:reproducibility}

This appendix records the retrieval, screening, and extraction settings used in the archived search run. The search was conducted programmatically on 26 July 2026 using the Semantic Scholar API as the primary retrieval source and covered publications from 2023 through July 2026. Table~\ref{tab:query-families} lists the query families and their task-specific terms. These were combined with LM-related expressions, such as "large language model", LLM, language/LLM agent, and agentic AI, as well as relevant cloud--edge--IoT system-context terms. The placement/offloading and resource management queries also used \texttt{-diffusion -GAN -CVAE -VAE} to reduce retrieval of generative AI studies without language-model involvement. The retrieved records totaled 1,808 before deduplication and 1,443 after deduplication.

\begin{table}[h]
\caption{Query families for corpus retrieval.}
\label{tab:query-families}
\centering
\footnotesize
\setlength{\tabcolsep}{4pt}
\renewcommand{\arraystretch}{1.05}
\begin{tabularx}{\textwidth}{
    >{\raggedright\arraybackslash}p{0.19\textwidth}
    >{\raggedright\arraybackslash}p{0.12\textwidth}
    >{\raggedright\arraybackslash}X}
\toprule
\textbf{Query family} & \textbf{Type} & \textbf{Task-specific terms} \\
\midrule
Orchestration core & Primary &
resource/service/network/policy orchestration; orchestrator; zero-touch management \\

Scheduling & Primary &
task, workflow, resource, or job scheduling \\

Placement/offloading & Primary &
service/function/container placement; task/computation offloading \\

Autoscaling & Primary &
autoscaling; SLA management; load balancing \\

Resource management & Primary &
resource allocation; resource management \\

Fault remediation & Primary &
fault diagnosis; RCA; remediation; self-healing; incident response; corrective action \\

Intent/policy & Primary &
intent-driven/based; intent translation; policy generation/orchestration \\

Agentic zero-touch & Primary &
zero-touch; autonomous network/orchestration; network/service orchestration \\

Execution safety & Supplementary &
verification; validation; closed-loop/feedback loop; execution or corrective action \\

Anomaly-to-remediation & Supplementary &
anomaly detection combined with RCA, remediation, troubleshooting, repair, or incident response \\
\bottomrule
\end{tabularx}
\end{table}

The deterministic desk screening used fixed metadata and lexical criteria rather than LM-generated judgments. Records were screened for the publication period, evidence of a language-model component, a systems or operations context, and relevance to the operational scope defined in Section~\ref{section/survey-methodology}. Records in which the LM was merely a workload being trained, served, deployed, or scheduled were treated as out-of-scope signals. Missing metadata, including a missing abstract, was not by itself used as an automatic exclusion criterion. The resulting 1,200 records entered semantic screening.

\begin{table}[h]
\caption{LM-assisted screening and extraction settings.}
\label{tab:screening-settings}
\centering
\footnotesize
\setlength{\tabcolsep}{3.5pt}
\renewcommand{\arraystretch}{1.0}
\begin{tabularx}{\columnwidth}{
    >{\raggedright\arraybackslash}p{0.29\columnwidth}
    >{\raggedright\arraybackslash}X}
\toprule
\textbf{Item} & \textbf{Setting} \\
\midrule
Screening model &
\texttt{deepseek-v4-flash(preview)} \\

Assessments &
Two assessments using distinct prompts with the same model; the second did not receive the first decision \\

Generation &
\texttt{temperature=0}; structured JSON output \\

Input &
Title, year, venue, publication type, fields of study, and abstract \\

Decision labels &
\texttt{primary\_include}, \texttt{related\_work},
\texttt{exclude}, \texttt{uncertain} \\

Required evidence &
Quoted LM-role and orchestration evidence from the supplied title or abstract \\

Strong candidate &
Both assessments returned \texttt{primary\_include}, each with confidence $\geq 0.80$, valid supporting evidence, and no unresolved publication-type ambiguity \\

Full-text extraction &
RQ extraction used \texttt{deepseek-v4-flash}; substantive fields
were associated with short source spans and source locations \\
\bottomrule
\end{tabularx}
\end{table}

The two screening prompts applied the same scope criteria but elicited separate assessments of each record. Each assessment assigned an operational-relevance label, while strong-candidate decisions required supporting textual evidence. The prompts distinguished substantive LM participation in orchestration from incidental mentions or LM-as-workload studies, allowing uncertain classifications when the available evidence was insufficient. The 421 records satisfying the strong-candidate rule subsequently underwent manual full-text eligibility assessment as described in Section~\ref{section/survey-methodology}.

For the comparative synthesis, LM-assisted extraction was used to identify and structure relevant full-text evidence according to the SPACE dimensions. The extracted records supported cross-study comparison, while evidence reported in the manuscript, figures, and comparative tables was manually verified against the corresponding primary studies. Ambiguous classifications were resolved through author discussion.

%For the comparative synthesis, LM-assisted extraction was used to locate and structure relevant full-text evidence. Extracted records served as an indexing and comparison aid; evidence carried into the manuscript, figures, and comparative tables was checked against the corresponding primary-study text.

\begin{table}[H]
\caption{Operational study-selection criteria.}
\label{tab:selection-criteria}
\centering
\footnotesize
\setlength{\tabcolsep}{3.5pt}
\renewcommand{\arraystretch}{1.06}
\begin{tabularx}{\columnwidth}{
    >{\raggedright\arraybackslash}p{0.27\columnwidth}
    >{\raggedright\arraybackslash}X}
\toprule
\textbf{Criterion} & \textbf{Operational rule} \\
\midrule
Publication period &
Publications from 2023 through July 2026. \\

Model role &
An LM or language agent must substantively contribute to planning, policy generation, decision-making, diagnosis, remediation, execution, or another identifiable orchestration function. Mere LM mentions, or LM-as-workload studies, are excluded. \\

System context &
Cloud, edge, IoT, cloud-native, distributed, or mobile-network infrastructure, including Kubernetes, microservices, serverless, MEC, O-RAN, 5G/6G, and zero-touch environments. \\

Operational relevance &
The study must address an identifiable orchestration or operational process, including scheduling, placement/offloading, scaling, resource management, intent/policy translation, diagnosis, incident response, remediation, or self-healing. \\

Study evidence &
Concrete methods, systems, prototypes, implementations, experiments, evaluations, or case studies with sufficient information to establish the LM's operational role. \\
\bottomrule
\end{tabularx}
\end{table}